# Combined Routing Protocol (CRP) for ad hoc networks: Combining strengths of location-based and AODV-based schemes.

Authors: Anton Sergeev, Victor Minchenkov, Aleksei Soldatov, Yaroslav Mazikov

The work proposes a new Combined Routing Protocol (CRP) for ad hoc networks [1-4 that combines the benefits and annihilates the shortcomings of two well-known on-demand routing protocols in ad hoc networks: AODV [5-7] (with provides high probability of successful discovering and maintaining a reliable route) and GPSR [9-10] (with fast on-the-fly transmission based on the geographical coordinates of the destination node). The main idea of the new routing scheme applied in CRP is to use AODV protocol as a solution to the "perimeter problem" of GPSR. And vice versa we apply GPSR for moving the starting point of the AODV route discovering closer to the destination point, decreasing the number of hops and route building time, making the resultant route more stable.

As the key result we see decreasing of the average packet delivery time in ad hoc networks with is extremely important for latency-critical applications, such as video streaming or command traffic [11].

## Routing in ad hoc networks

Ad hoc networks can be described as wireless multi-switching networks without prior infrastructure, with topology changing due to node mobility. Each node in the ad hoc network operates simultaneously like a terminal system and a router [1-4]. Ad hoc network applications are communication between mobile objects/devices, data exchange and accumulation operations in an unfriendly environment. In ad hoc networks, nodes are not aware the networks' topology. Instead, they must discover it Routing protocols in a wireless ad hoc network can be divided into 2 large groups[2-4]:

- protocols that use information about the network topology
- protocols that use information about the geographical location of nodes in the network.

For topology-based protocols, a new device usually announces its presence and listens for announcements broadcast by its neighbors. Each node learns about others nearby and how to reach them, and may announce that it too can reach them. Topology-based routing protocols use information about the presence or absence of communication channels between nodes. The process of operation of these protocols can be divided into 3 main stages: discover and build the route, transmit messages between nodes in accordance with the route, maintaining the route. Examples of topology-based routing Dynamic Source Routing (DSR) [8] or Ad hoc On-Demand Distance Vector (AODV) [5-7].

The location-based routing (e.g. Greedy Perimeter Stateless Routing, GPSR [9-10]) allows routers to be nearly stateless, and requires propagation of topology information for only a single hop: each node need only know its neighbors 'positions.

Next, we will briefly describe some of the most prominent and well-known representatives of both types of protocols and propose a solution that compensates for the disadvantages of both classes, while maintaining the advantages.

**Protocols to be combined: short description, problems, advantages of AODV and GPSR**

**AODV** is a reactive or on-demand routing protocol which means a route between two nodes will be determined only when there is data to be transmitted [5-6].

For each object in the network, the routing tables contain information about the next node on the way to the destination node, timestamps, and service information. Routes are supported only if they are used. To avoid looping and confirm that the route is fresh really sequential numbers AODV uses sequential numbers.

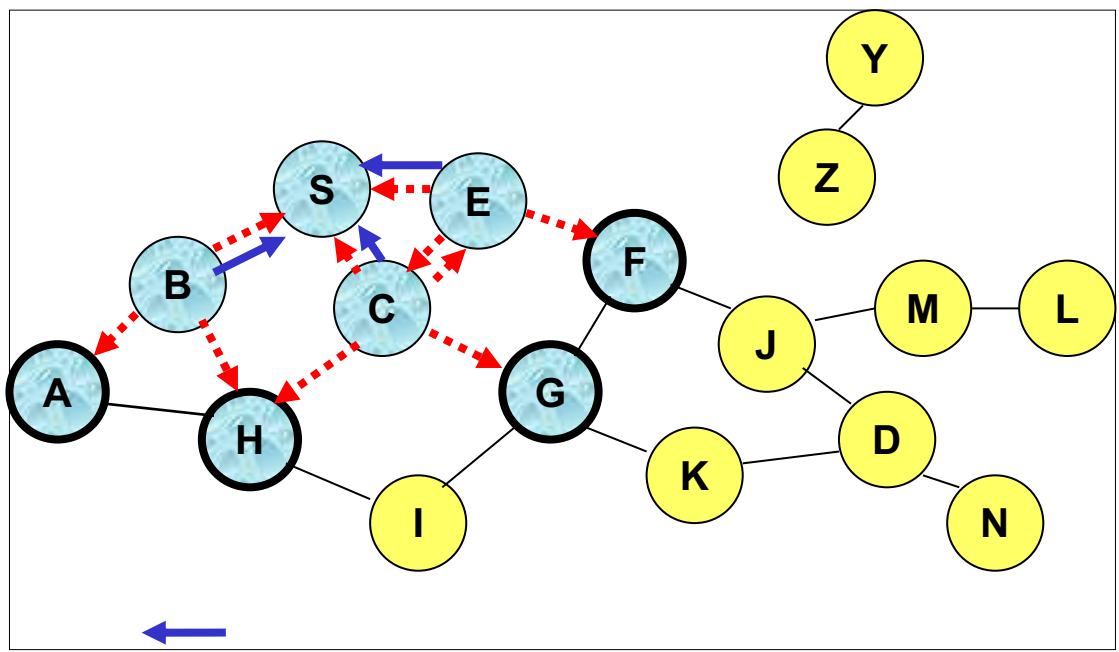

Picture 1 Operation of the AODV protocol when discovering a new route (from node A to D): broadcasting of RREQ messages and establishing reverse connections [19]

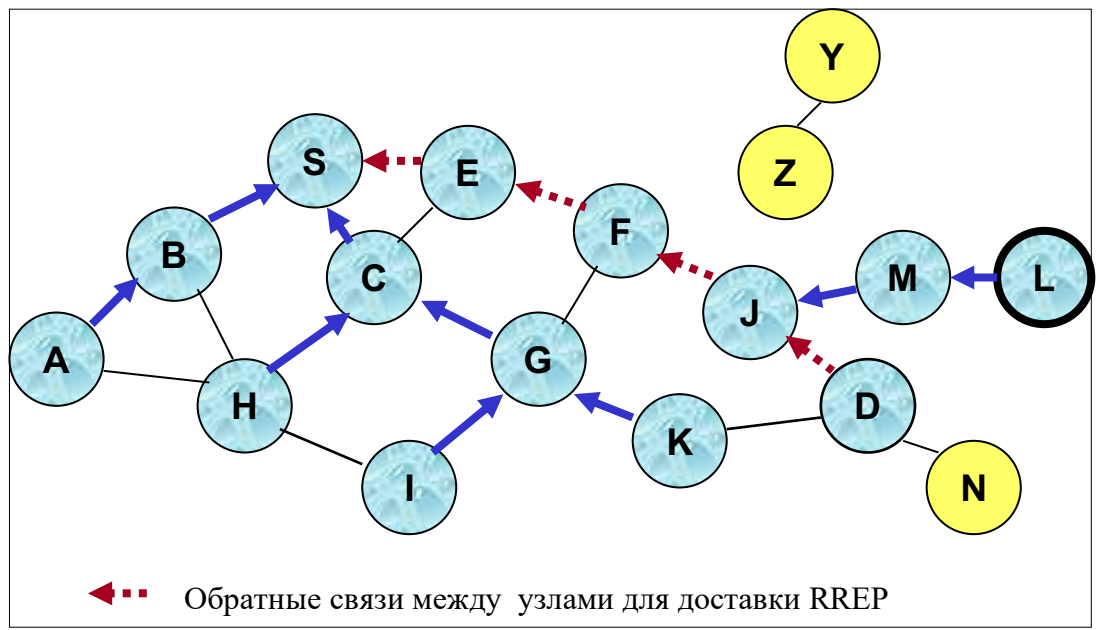

Picture 2 Operation of the AODV protocol when a new route is discovered: RREP delivery via reverse connection and setting a new path from the sender (A) to the destination node (D) [19]

AODV uses the following types of packages:

- Welcome messages
- Communication Route Request (RREQ)
- Communication Route Response (RREP)
- Communication Route Error (RERR)

Welcome messages are periodic local radio broadcasts informing each mobile node in its neighbors. The use of welcome messages is optional. Nodes can also monitor data packet retransmissions to ensure that the next switch is still within reach.

When a node has a data packet to transmit and does not know the path to the destination, it transmits an RREQ broadcast message to all its neighbors. RREQ messages will be propagated through the network in all directions until they reach a destination or node with a sufficiently fresh route. During the RREQ redirection process, intermediate nodes make entries in their routing tables for the neighboring node from which the first copy of the broadcast packet was received (Pic. 1).

As soon as the RREQ reaches a destination or intermediate node with a sufficiently fresh route, the destination/intermediate node responds by unidirectionally transmitting the RREP packet back towards the neighbor from which the RREQ was received. The RREP is routed back to the source node along the return path, establishing a route on the intermediate nodes for the subsequent transmission of data packets (Pic 2). Due to the reverse path, AODV supports only symmetric communication channels.

The found communication direction is maintained by the source node as long as it is needed. If the source node moves during an active communication session, the route exploration can be initialized again to establish a new communication direction to the destination. When the destination or intermediate nodes change their position, a RERR message is sent to the corresponding source nodes. After the source node receives the RERR, the route exploration can be reinitialized, if necessary.

The main advantages of AODV protocol:

- Communication directions are supported only between the nodes that need to communicate. This reduces the overhead of maintaining a communication route.
- There is no need to include communication directions in the packet headers. Therefore, AODV is more scalable than, for example, DSR.

The main AODV disadvantages:

- The main problem for AODV, as for most reactive protocols, is the situation of breaking an already laid route. Every time an intermediate or final node leaves the radio coverage area of its neighbors in a minibus, a gap occurs in it. A RERR message is sent to the affected nodes, which triggers, if necessary, a comprehensive procedure for restoring the severed route. In this case, the search process is reinitiated. The probability of route disruption increases with increasing node mobility and route length, and therefore with increasing network size.
- Does not support unidirectional communication channels.
- Does not support multiple routes to the destination.

**Greedy Perimeter Stateless Routing (GPSR)**

**GPSR** [9] refers to position-based routing protocols. When using the "greedy" algorithm, the sending node includes information about the geographical location of the receiving node in the packet header. As a result, for the node relaying the packet, a "greedy" solution for transmitting the message to the geographically nearest node will be optimal locally.

According to this protocol, the source node includes the coordinates of the destination node in each transmitted packet. [2],[6],[7]. Intermediate nodes decide on the further direction of transmission (select the next node) based on the coordinates of the destination node contained in the packet and the coordinates of their neighbors (the neighbor closest to the destination node is selected in terms of geographical distance). If there are no nodes closer to the recipient (the main problem of this protocol) than the current node, the perimeter mode is used, based on bypassing such an empty space according to the right-hand rule.

A situation where the intersection (Figure 3) of the decoding radius of node x and the circle around node D (this may be the recipient or the intermediate node) It does not contain nodes and is called the "perimeter problem". From the position of node x, the intersection area is called an empty zone (void). Thus, node x has no neighbors closer to the recipient than itself.

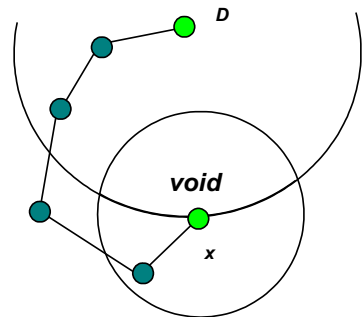

Picture 3. Empty area on node x during package delivery to node D.

The algorithm of "greedy" routing reaches a "local maximum" from which it cannot exit. At the same time, node x tries to recover and bypass the empty zone. Obviously, the route constructed in this way does not contain nodes from an empty zone (otherwise, the "greedy" routing algorithm would work).

In case of a "perimeter problem", the GPSR switches to perimeter bypass mode. If we turn to the representation of a wireless network in the form of a graph, where the edges correspond to the connections between the nodes, then the route to the recipient obtained by bypassing the empty zone will pass along the edges bordering this area. GPSR forwards the packet along the outer edges of the planar subgraph according to the right-hand rule (transmitting the packet along the next edge counterclockwise, forming the smallest angle with the edge along which the packet arrived at the current node), gradually bringing it closer to the recipient.

Theoretically, this mode guarantees finding the path to the destination node, bypassing the empty zone, if it exists. The "perimeter problem" is the main one for all schemes that use a "greedy" routing algorithm and worsens the already <u>serious shortcomings of the GPSR protocol</u>:

- Routes are becoming even more suboptimal and longer
- Due to inefficient routes, continuous node movement, and packet lifetime (TTL) restrictions, the probability of packet dropping and deletion at intermediate nodes increases.
- The effectiveness of the protocol depends on the network graph planarization algorithm.
- Packet Delivery Success Rate strongly depends on every node's immediate neighbors and their mutual geographical position.
- In addition, the receiving node may change its location during the data transfer process. And GPSR does not have the means to update this information during route setup, which, again, leads to even longer routes and package deletion due to limitations on its lifetime. That is, GPSR essentially "misses" by relying on data from a geographic information delivery system that is not fully consistent with the routing algorithm.
- The effectiveness of the protocol also decreases due to the distance effect, an effect where it seems that the nearest nodes are moving at a higher angular velocity than the farthest nodes. For this reason, the difficulty of finding the right and optimal path increases as the distance to the recipient decreases.

At the same time there are some <u>significant advantages of GPSR protocol</u>:

- It's fast. There no stage of route discovery – the GPSR makes greedy forwarding decisions using only information about a router's immediate neighbors in the network topology.
- There are no stage of initial route discovery that greatly decreases packets' transmission delay for short routes.
- According to the protocol's authors «greedy forwarding's great advantage is its reliance only on knowledge of the forwarding node's immediate neighbors. The state required is negligible, and dependent on the density of nodes in the wireless network, not the total number of destinations in the network.1 On networks where multi-hop routing is useful, the number of neighbors within a node's radio range must be substantially less than the total number of nodes in the network».
- Much smaller Routing Protocol Overhead in compassion both with DSR and AODV

**Combined Routing Protocol (CRP)**

An effective combination of various routing schemes (reactive and proactive, operating on the request of topological and location-based protocols) – is one of the open problems of modern ad hoc networks.

The main goal in the design and research of the hybrid schemes proposed in this paper was to try to combine the advantages of two protocols: GPSR and AODV. The main idea of the new routing schemes is to use the AODV protocol as a solution to the "perimeter problem" of GPSR. The formal description of the combined routing protocol is given below:

**IF possible**, *we use GPSR in the "Greedy" mode*: on-the-fly transmission of packets to the nearest node, geographically closer to the recipient.

**ELSE**, *a simplified version of AODV is used*:

- without a route recovery procedure in case of a break,
- using a callback from the MAC level instead of periodic hello messages to exchange data on the position of neighbors. A route is discovered from the current node to the recipient (instead of the perimeter bypass mode), followed by data transmission along this route.

The main idea of the proposed scheme is to deliver the data packet closer to the recipient using the GPSR mode, and in case of empty zones, use the AODV mode to find the way to the recipient.(Picture 4).

**Combined GDSR+AODV Routing …**

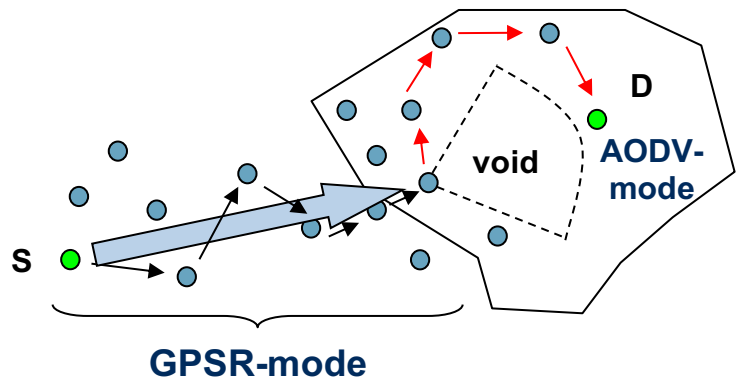

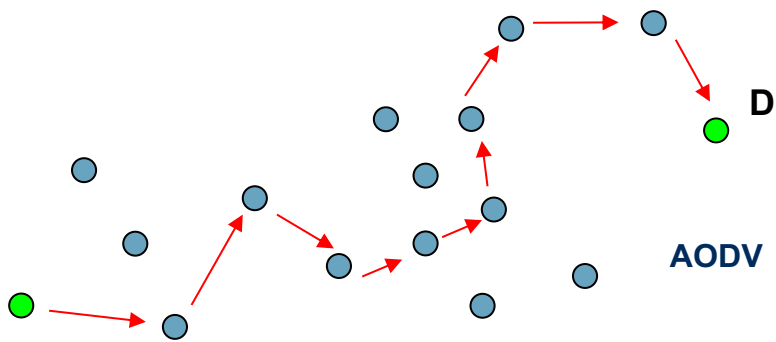

Picture 4. AODV as a solution to the "perimeter problem" of GPSR

## Planning an experimental protocol comparison depending on the topology, load, and mobility level of nodes in the network

In the full version of the article, it is planned to publish the results of modeling and performance analysis (throughput, delay, network overhead) of new hybrid routing schemes (CRP v.1, CRP

v.2, Caching GPSR) in comparison with the basic protocols on the basis of which they were created (AODV, GPSR). The comparison is made depending on the network load.

The work analyzes the simulation results and evaluates the performance of new proposed combined routing protocols (CRP v1, v.2) and schemes using topology data (AODV, DSR) or the geographical location of nodes in the network (GPSR). The simulation was performed for wireless networks with different topologies, traffic scenarios, load levels, and mobility.

The process of modeling and analyzing the performance of the proposed routing schemes in comparison with existing protocols is divided into 2 stages:

At the first stage (subsection 4.2.1), AODV, GPSR and new hybrid protocols (CRP v.1, CRP v.2, Caching GPSR) are compared depending on network load. A network of 30 nodes with variable CBR traffic is used, increasing in the range from 25 to 1 packet/second. For each type of scenario and all levels of network dynamics, 10 scenarios with a duration of 500 seconds are generated.

The simulation parameters are presented in the table Table 1.

| Number of nodes | Size of mobility area | Shutdown time, sec | Packet size, bytes | Traffic (pause between packets, sec) | Time of experiment, sec |
|---|---|---|---|---|---|
| 30 | 1000 m × 1000 m | 40 | 512-bytes | 0,25 CBR | 500 сек. |

Table 1. Network scenario parameters for investigating
the impact of network load on protocol performance

Then, in the second stage (subsection 4.2.2), the DSR and GPSR are compared with the first hybrid protocol (CRP v.1) depending on the steppe mobility of nodes. The selected mobility levels in this study are determined by pauses in node movement during: 0 seconds, 10 seconds, 20 seconds, and 40 seconds. The average speed is 20 m/s. 4 data packets are generated per second. For each type of scenario and all levels of network dynamics, 10 scenarios are generated with a duration of 500 seconds (120 scenarios in total).

| Number of nodes | Area | Density of nodes | Number of data streams |
|---|---|---|---|
| 30 | 1000m X 1000m | 1 узел / 33333 м$^2$ | 20 |
| 50 | 1000m X 1000m | 1 узел / 20000 м$^2$ | 20 |
| 100 | 1000m X 1000m | 1 узел / 10000 м$^2$ | 20 |

Table 2. Types of simulation scenarios

Next, for each protocol, the average values of the values indicated above are calculated for all three types of scenarios for all levels of network dynamics.

Performance evaluation, protocol analysis, and comparison are performed based on the following values:
- o average delay in data packet transmission;
- o protocol overhead;
- o bandwidth.

<u>Throughput</u> is measured as the ratio of the number of data packets that have reached the destination node to the total number of packets transmitted.

<u>Transmission delay</u> is defined as the time interval that elapses from the moment a data packet is generated by the sending node to the moment it is received by the receiving node.

<u>Overhead</u> here is the total number of service and data packets transmitted to the network by all nodes during the simulation. The number of transmissions of all packets is calculated: redirection is considered as another transmission; broadcast transmission is considered as one transmission.

## Acknowledgements

The research was carried out within the framework of the HSE Fundamental Research Program and the Strategic Project "Digital Transformation: Technologies, Effects, Efficiency" of the HSE. The study was carried out using the HSE supercomputer complex.

## Sources